\documentclass[prl,twocolumn,preprintnumbers,superscriptaddress,amsmath,amssymb,floatfix,showpacs]{revtex4-1}
\usepackage{amsmath,times,graphicx,bm,nicefrac,tikz,textcomp,amssymb,mathptmx}

\newcommand{\chirp}{\mathcal{C}}
\newcommand{\Felix}[1]{\textcolor{black}{#1}}

\begin{document}

\allowdisplaybreaks

\title{Chirped standing wave acceleration of ions with intense lasers}

\author{F. Mackenroth}
\email{felix.mackenroth@chalmers.se}
\affiliation{Department of Physics, Chalmers University of Technology, SE-41296 G\"oteborg, Sweden}

\author{A. Gonoskov}
\affiliation{Department of Physics, Chalmers University of Technology, SE-41296 G\"oteborg, Sweden}
\affiliation{Institute of Applied Physics, Russian Academy of Sciences, Nizhny Novgorod 603950, Russia}
\affiliation{Lobachevsky State University of Nizhni Novgorod, Nizhny Novgorod 603950, Russia}

\author{M. Marklund}
\affiliation{Department of Physics, Chalmers University of Technology, SE-41296 G\"oteborg, Sweden}

\date{\today}
\begin{abstract}
We propose a novel mechanism for ion acceleration based on the guided motion of electrons from a thin layer. The electron motion is locked to the moving nodes of a standing wave formed by a chirped laser pulse reflected from a mirror behind the layer. This provides a stable longitudinal field of charge separation, thus giving rise to \textit{chirped standing wave acceleration} (CSWA) of the residual ions of the layer. We demonstrate, both analytically and numerically, that \Felix{stable proton beams, with energy spectra peaked around 100 MeV,} are feasible for pulse energies at the level of $10$ J. Moreover, a scaling law for higher laser intensities and layer densities is presented, indicating stable GeV-level energy gains of dense ion bunches, for soon-to-be available laser intensities.
\end{abstract}

\maketitle

\begin{figure}[htp]\centering
 \includegraphics[height=0.6\linewidth]{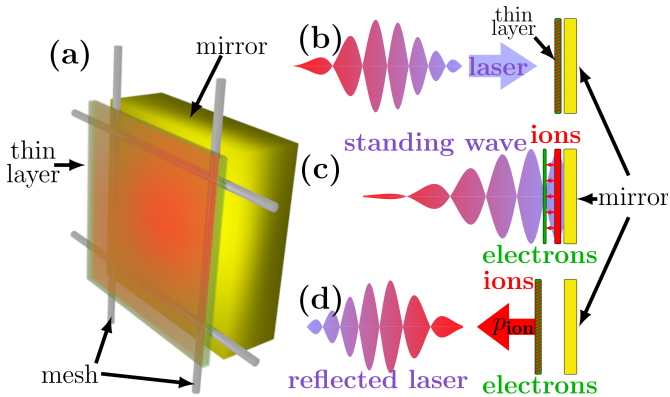}
 \caption{(color online) \Felix{(a): CSWA target: High density mirror and a thin layer fixed in a certain position in front by a $\mu$m thin mesh, leaving voids for the standing wave to form. (b): Chirped laser pulse impinges on the target. (c): Standing wave forms, locks and displaces the thin layer's electrons which drag the layer's ions. (d): Ions and electrons propagate away from the mirror.}}
 \label{fig:Setup}
\end{figure}

The acceleration of charged particles by intense lasers has become a cornerstone of plasma physics research \cite{Daido_etal_2012,Macchi_etal_2013}, \Felix{especially aiming at medical applications requiring stable proton beams of $100$-$200$ MeV energy \cite{SSBulanov_etal_2008a},} and there are several approaches \cite{Mackenroth_etal_2016}. In Target Normal Sheath Acceleration (TNSA) \cite{Roth_etal_2002,Mora_2003,Cowan_etal_2004,Passoni_etal_2010}, as well as Coulomb explosion of clusters \cite{Ditmire_etal_1997} or specially designed targets \cite{Esirkepov_etal_2002,SSBulanov_etal_2008b}, the plasma electrons are heated locally and the remaining ions accelerated by electrostatic fields arising due to the electron cloud's expansion or the ions' repulsion, respectively. The robustness of this process makes it experimentally accessible. But the heating also deprives one of control of the energy transformation making it difficult to deliver the laser energy to a certain range of ion energies. This inevitably results in a broad ion spectrum, a comparatively low efficiency and a rather unfavorable scaling law for the ion energy as a function of the laser pulse intensity \cite{Daido_etal_2012,Macchi_etal_2013}. Collisionless shock acceleration \cite{Silva_etal_2004,Haberberger_etal_2012} can provide monoenergetic ion bunches, however, typically involves many energy transformation steps also yielding low efficiency and a small number of accelerated ions. Hole boring \cite{Schlegel_etal_2009}, light sail, or laser piston, \cite{Esirkepov_etal_2004,Bulanov_etal_2010,Henig_etal_2009,Kar_etal_2012} imply a more direct and thus controllable energy transfer, yielding higher efficiency as well as more promising spectral properties and scaling laws. However, the balance between the plasma fields and the light pressure introduces an interface susceptible to instabilities \Felix{\cite{Pegoraro_Bulanov_2007,Khudik_etal_2014,Sgattoni_etal_2015,Palmer_etal_2012}}, which strongly limits the acceleration and can even make it experimentally unfeasible.

In this Letter we propose a new basic approach, which provides an opportunity to overcome the outlined obstacles. \Felix{The approach relies on placing a thin plasma layer close to a reflecting mirror (s.~Fig.~\ref{fig:Setup}~(a)) and locking its electrons between a standing wave's (electric) field antinodes to move them controllably.} This leads to the creation of a capacitor-like electric field due to charge separation between the shifted electrons and the parent ions, the latter being less affected by the ponderomotive force but still being dragged by the electron layer. As the process involves ponderomotive confinement of the electrons from both sides, the locking mechanism does not introduce any interface susceptible to instabilities. The necessary control can be achieved by reflecting a chirped laser pulse from a mirror placed behind the layer. The well controlled change of the laser's wavelength leads to a changing position of the electric field nodes, similar to studies of ion acceleration in vacuum \cite{Peano_etal_2008a,Peano_etal_2008b}. Consequently, the motion of the locked electrons can be coordinated with the ion acceleration in the charge separation field, making the ions follow the locked electrons for a long distance. We label this concept \textit{chirped standing wave acceleration} (CSWA). In principle, it only relies on a tunable laser chirp and a large bandwidth. This is becoming accessible in the ultra-high intensity regime, \Felix{with several J of laser energy available at a relative bandwidth of $60\%$ \cite{PFS}, and further development towards higher laser energies at large bandwidths planned \cite{Zepf_PC}.} 

We model the incident laser field as a circularly polarized plane wave of electric field amplitude $E_0$ and angular frequency $\omega_0$ depending on space-time only via the invariant phase $\eta=t-x/c$ where $c$ is the speed of light and $x$ ($p$) is the coordinate (momentum) along the laser's propagation direction. Upon reflection from a mirror placed at $x=0$ the laser will form a standing wave $E_\text{tot}(t,x) = E(\eta) - E(\eta_+)$, where $\eta_+=t+x/c$. \Felix{Since the heated mirror's sheath field spatially only extends over a Debye length $\lambda_D\ll \lambda_0/2$ it will not affect a plasma layer of areal density $\sigma$, placed at the standing wave's first node.} The layer electrons (mass and charge $m_e$ and $e$, respectively) will be trapped by the ponderomotive force $F_\text{pond} =- e^2\left<E_\text{tot}^2(t,x)\right>/2m_e\omega_0^2 $, where the brackets indicate a temporal average. The ions, on the other hand, are less affected by this force. In contrast, they experience the electron layer's electrostatic force which varies from $F_C=0$ to \Felix{$F_C^\text{max}=4\pi e^2\sigma$}, depending on the ions' position. \Felix{To estimate achievable peak ion energies we use $F_C^\text{max}$. We determine the layer's areal density $\sigma$ by requiring that, in} order to place a layer of particles in front of the mirror and still allow a standing wave to form upon incidence of a laser wave, the thin layer needs to transmit radiation in the relevant frequency regime. The electrons will, however, upon incidence of the laser, be accelerated to form a current $j$ emitting radiation canceling the incoming one. Since the current per unit area is limited by $j_{\text{max}} = e \sigma c$ intensities $I>I_{\text{th}} = c/\left(4\pi\right)\times\left(2\pi j_{\text{max}}/c\right)^2 = \pi c e^2 \sigma^2$ cannot be canceled and are transmitted through the layer. This is a simple particular case of relativistic self-induced transparency (RSIT) \cite{Vshivkov_etal_1998,Macchi_etal_2009}, as applied in the break-out afterburner \cite{Yin_etal_2006,Hegelich_etal_2006,Jung_etal_2013a,Jung_etal_2013b,Hegelich_etal_2013}. In dimensionless units, a radiation field of amplitude $a_0 = eE_0/m_e c\omega_0 $ is transmitted trough a plasma layer with areal density below the threshold $\sigma_0 = \sigma_\text{cr}a_0/\pi$, where $\sigma_\text{cr}=2\pi c n_\text{cr}/\omega_0 $ with the critical plasma density $n_\text{cr} = m_e\omega_0^2/4\pi e^2$. The Coulomb attraction to the residual ions of the layer dominates over the light pressure force $F_l = (2/c)I$, until the threshold value $I_{\text{th}}$ is reached. Thus, almost immediately after starting traveling beyond the layer, the electrons get trapped by the newly established standing wave, making this scheme stable against the loss of electrons due to radiation pressure into the mirror. Unlike in the staged ion acceleration scheme \cite{Gonoskov_etal_2009}, the electrons are here locked in the resulting standing wave.

\begin{figure}[t]\centering
 \includegraphics[width=\linewidth]{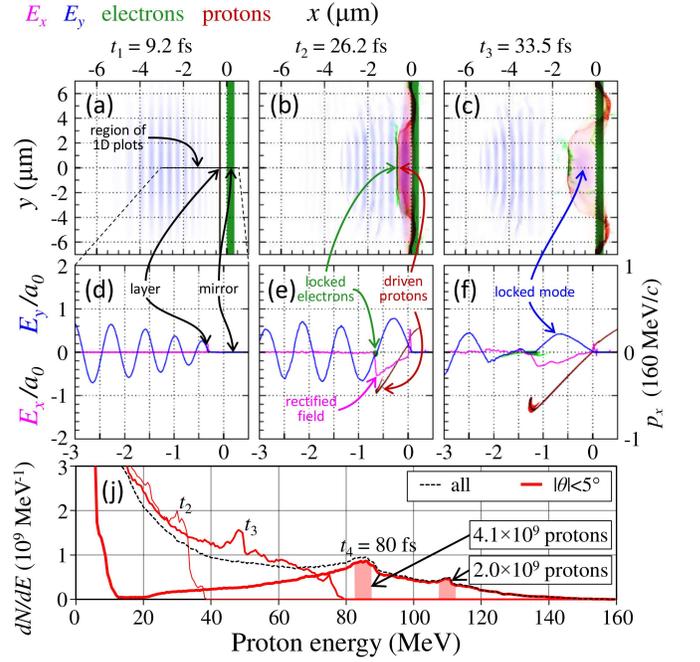}
 \caption{(color online) 2D PIC simulation for a laser energy $\varepsilon_0=30$~J , bandwidth $\Delta\omega_0=0.5\omega_0$, chirp $\chirp=-3.5$. (a)-(c): Transverse field $E_y$ (blue), electrons (green) and protons (red) as functions of 2D coordinates. (d)-(f): 1D cut additionally showing longitudinal field component $E_x$ (magenta) and particle distributions in phase space $x$-$p_x$. Three time instants shown: Before the layer starts transmitting the incident radiation [(a),(d)], during CSWA with a standing wave formed in the laser's reflection [(b),(e)] or by a mode of radiation locked between the layer and the mirror [(c),(f)]. (j): \Felix{Spectra of protons from the thin plasma layer (effectively 3D) at times $t_2$, $t_3$ and a later time $t_4$ propagating within a forward cone of $10^\circ$ opening angle (red lines) or in an arbitrary direction at $t_4$ (black dashed line).}}
 \label{fig:PhaseSpace}
\end{figure}

The amplitude $a_0$ of the laser is a function of time and the laser will be transmitted through the thin layer only during the time interval $\tau_\text{acc}$ while it is relativistically transparent. We assume the electron layer to be confined by the laser field throughout this whole time interval whence the ions will be approximately dragged by $F_C^\text{max}$ for a time $\tau_\text{acc}$. The ions' final momentum then is
\begin{align}\label{Eq:Momentum}
 p_\text{ion} &= 4\pi e^2\sigma \tau_\text{acc}.
\end{align}
The ion acceleration thus has two extreme cases where either the charge density vanishes, allowing for an immediate break-through of the electric field, or where the layer is so thick that the laser is not able to break through. In both cases the ions will not be significantly accelerated. Thus, there exists an optimal value for $\tau_\text{acc}$. To estimate this, we neglect the oscillating phase structure  of the field and assume that the laser pulse has a Gaussian temporal shape with some bandwidth $\Delta\omega$. The amplitude at the time of break-through is then given by $a_{0}(\tau_\text{acc}/2 ) = a_0 \text{exp}[-(\Delta\omega \tau_\text{acc}/2)^2]$. Inserting the optimal areal density we find the corresponding areal density
\begin{align} \label{Eq:OptimumDensity}
 \sigma = \frac{2c n_\text{cr}a_0}{\omega_0} \text{e}^{-\left(\Delta\omega \frac{\tau_\text{acc}}2\right)^2},
\end{align}
and the final ion momentum is given by
\begin{align}
 p_\text{ion} &= 2c m_e\omega_0a_0 \left[\text{e}^{-\left(\Delta\omega \frac{\tau_\text{acc}}2\right)^2} \tau_\text{acc}\right]. \label{Eq:IonMomentum}
\end{align}
Only the term in brackets depends on the accelerating time. The maximum of this expression is found at $\tau_\text{acc}^\text{opt} = \sqrt2/\Delta\omega$. We note that the above equation is already optimized for a given set of layer parameters, since the layer's areal charge density is chosen such that relativistic transparency, allowing the pulse to break through the layer, sets in at the time $t=-\tau_\text{acc}/2$.
Eq.~(\ref{Eq:IonMomentum}) describes a particle of charge $e$ accelerated in a constant field $E_0\,\text{exp}(-1/2)$ over a time span $\tau_\text{acc}^\text{opt}$. One can thus view the present scheme as a highly efficient field rectifier which turns the laser pulse's transversal into an accelerating longitudinal field of approximately the same amplitude, as also confirmed by numerical simulations (s.~Fig.~\ref{fig:PhaseSpace}~(e)). To shift the trapped electron layer controllably, one can use a chirped laser pulse, continuously changing its wavelength, and hence also the position of the field nodes. Using a chirp has also been suggested to control other ion acceleration schemes \cite{Peano_etal_2008a,Peano_etal_2008b,Peano_etal_2009,Galow_etal_2011,Popp_etal_2010,Pathak_etal_2012,Vosoughian_etal_2015}. However, whereas these previous works in the literature build on the model of a linear pulse chirp, we here instead use a model of a laser pulse chirp beyond the linear approximation, in close analogy to the model of a chirped pulse amplifier \cite{Mourou_etal_2006,FundamentalsPhotonics_2013}. The model is based on a frequency dependent phase shift of the field's frequency components (s.~Supplementary material). According to this model a chirped plane wave laser pulse with Gaussian envelope is given by (s.~also \cite{FundamentalsPhotonics_2013})
\begin{align}\label{Eq:ChirpedField}
 E\left(\eta\right) &= \frac{E_0}{\left(1+\chirp^2\right)^{-\nicefrac14}}\text{e}^{-\left(\Delta\omega_0\left(\chirp\right) \eta\right)^2 +i \Sigma\left(\eta\right)}\\
 \Sigma\left(\eta\right) &= \omega_0 \eta + \chirp\left[\left(\Delta\omega_0\left(\chirp\right)\eta\right)^2  + \frac{2\omega_0^2 \log2 }{\Delta\omega_0^2} \right]+\frac{\text{atg}\chirp}{2} \nonumber,
\end{align}
with the bandwidth $\Delta \omega_0$ (connected to the FWHM pulse duration $\Delta\tau= 4\log2\sqrt{1+\chirp^2}/\Delta\omega_0\Felix{\approx1.2 \text{ fs} \times\sqrt{1+\chirp^2}\omega_0/\Delta \omega_0}$), $\chirp$ a dimensionless parameter quantifying the chirp strength and $\Delta\omega_0\left(\chirp\right) = \Delta\omega_0/\sqrt{8\log2\left(1+\chirp^2\right)}$. The two perpendicular field components of a circularly polarized laser pulse are then $\bm{E} = \left(\text{Re}\left(E\left(\eta\right)\right),\text{Im}\left(E\left(\eta\right)\right)\right)/\sqrt{2}$. The frequency changes as a function of $\eta$ according to $\omega(\eta) = \omega_0 + 2 \chirp\Delta\omega_0^2\left(\chirp\right)\eta$, showing that we go beyond the linear chirp approximation. The standing laser wave formed upon reflecting the field (\ref{Eq:ChirpedField}) from a mirror fulfills the energy balance $E_\text{tot}^2(t\to-\infty,x\to-\infty)=E_\text{tot}^2(t\to+\infty,x\to-\infty)$, highlighting that no radiation pressure is involved in the acceleration. Neglecting the temporal envelope, the standing wave has \Felix{its $n^\text{th}$ node} in the negative half-space at the purely time-dependent position
\begin{align}\label{Eq:NodePosition}
\Felix{x_\text{node}(t) = -n\frac{\pi c}{\omega(t)}.}
\end{align}
This node will move at a speed 
\begin{align}
 \Felix{v_\text{node}(t) = 2\pi n\,c \frac{\Delta\omega_0^2\left(\chirp\right)}{\omega^2(t)}\chirp}, \label{Eq:NodeVelocity}
\end{align}
moving the electrons locked to it. \Felix{Apparently at large $n$ the nodes can travel with speeds exceeding $c$.} Equating the ponderomotive force to $F_C^\text{max}$, we find the layer's equilibrium thickness
\begin{align}
  \Delta x = \frac{\sqrt2\pi c\omega_0\left(1+\chirp^2\right)^{\nicefrac34}}{a_0\omega^2(\eta)}.
\end{align}
Realizing that for a realistic pulse $\omega(\eta)\sim \omega_0$ (up to a factor of order unity) we see that for a relativistic field strength ($a_0\gg1$) the electron layer will be compressed by the ponderomotive force to a thickness $\Delta x \sim \lambda_0/a_0\ll \lambda_0$. Analogously, it can be shown that for any single particle separated further from the electron layer than twice the derived layer thickness, the laser's ponderomotive force dominates over the Coulomb attraction between the electron and ion layers, ensuring the stability of the suggested scheme throughout the whole duration of the laser pulse. Inserting the field model from Eq.~(\ref{Eq:ChirpedField}) into the optimized Eq.~(\ref{Eq:IonMomentum})  we obtain
\begin{align}\label{Eq:OptimalMomentum}
 p_\text{ion}^\text{opt} \approx 4m_ec\frac{\omega_0}{\Delta\omega_0}a_0\left(1+\chirp^2\right)^{\nicefrac14}.
\end{align}
The ions require a finite momentum even for $\chirp\equiv0$ as an artifact of modeling the accelerating field as constant over $\tau_\text{acc}$ in Eq.~(\ref{Eq:Momentum}). We thus apply Eq.~(\ref{Eq:OptimalMomentum}) only for $\left|\chirp\right|>1$. According to Eq.~(\ref{Eq:OptimalMomentum}) the ions' final momentum scales as $p_\text{ion}^\text{opt}\sim \left(1+\chirp^2\right)^{1/4}$, while the charge surface density scales as $\sigma \sim {\left(1+\chirp^2\right)^{-1/4}}$. The areal charge current density $j=\sigma p_\text{ion}^\text{opt}/m \sim a_0^2$, however, is independent of the chirp. We conclude that, provided one maintains the optimum surface density condition, the pulse chirp is a tunable parameter to trade maximum particle energies for number of accelerated particles. \Felix{Furthermore, it even allows to tune between high ion energies and a monoenergetic spectrum since for large chirps some of the ions outrun the locked electrons and are no longer accelerated. They start circulating around the locked electrons (s.~Fig.~\ref{fig:PhaseSpace}~(e,f)), compressing the protons' spectrum.} 
For nonrelativistic ions of mass $m_\text{ion}$, Eq.~(\ref{Eq:OptimalMomentum}) corresponds to a final ion velocity $v_\text{ion} \approx p^\text{opt}_\text{ion}/m_\text{ion}$. Since the ions need to be close to, but cannot overtake, the electron layer locked to the field nodes, we expect optimal ion acceleration if $v_\text{ion}\approx v_\text{node}(1.5\Delta \tau)$, where $1.5\Delta\tau$ is a suitable time scale for the ions to approach their final velocity. For $\left|\chirp\right|\gg1$ we then find the following chirp value for optimal ion acceleration
\begin{align}
 \chirp^\text{opt} \approx -\left[\frac{m_\text{ion}}{m_e}\frac{\pi}{16 \log2}\frac{\left(\frac{\Delta \omega_0}{\omega_0}\right)^3}{a_0 \left(1 - 1.5\left(\frac{\Delta \omega_0}{\omega_0}\right)\right)^2}\right]^{\nicefrac23}.\label{Eq:OptimalChirp}
\end{align}
The scaling $\chirp^\text{opt}\sim a_0^{-2/3}$ is due to that at higher pulse energies higher final ion velocities require faster node propagation.

The dominant higher-dimensional effect is dephasing: in the one-dimensional analysis one neglects the accelerating field's dependence on the distance between the proton and electron layers, effectively ending the acceleration once the layers are separated further than their transverse size. Since the layers' transverse size is, however, of the order the laser spot, i.e., several $\mu$m and the longitudinal motion is sub-wavelength, one expects dephasing to be negligible. Quantitatively, inserting the optimal areal charge density from Eq.~(\ref{Eq:OptimumDensity}) into the accelerating force $F_C^\text{max}$ we estimate the position of the accelerated ion layer $x_\text{ion}(t) = F_C/2m_\text{ion}t^2 + x_\text{ion}(0)$. Equating the resulting distance between the ion layer and the electrons, locked to the field nodes at $x_\text{node}(t)$, to the transverse spot size leads to a cubic equation in time, which we have investigated numerically (s.~Supplementary material). It showed that for all parameters studied in this work the time scale for dephasing is significantly longer than the pulse duration $\Delta\tau$, when the electrons are no longer driven by the standing wave and dephasing can no longer occur. We thus employ the presented one-dimensional analysis as a qualitative model of CSWA.

To test the analytical model we performed a series of numerical experiments. We note that due to the laser's circular polarization and circular focal spot there is no preferred transverse direction. Thus, \Felix{the formation of instabilities and the beam's divergence can be studied already in a 2D simulation.} Also, the total particle number in a 3D geometry can be deduced from a 2D simulation \Felix{resolving only the laser's propagation direction $x$ and one perpendicular coordinate $y$} via $N= 2\pi r (dN/dz)$, where $dN/dz$ is the particle density in the unresolved third coordinate and $r$ the distance from the laser axis \Felix{in $y$-direction. We performed a 2D particle-in-cell (PIC) simulation, using the code PICADOR \cite{PicadorTeam}, to demonstrate CSWA to efficiently produce an ion beam with low divergence which is stable against plasma instabilities} (s.~Supplementary video and Fig.~\ref{fig:PhaseSpace}). We used Eq.~(\ref{Eq:ChirpedField}) to model a circularly polarized laser pulse with central wavelength $\lambda_0 = 810$ nm under normal incidence, focused to a circular spot of $d_0=7.5\, \mu$m diameter \Felix{onto a thin layer of electrons and protons (mass $m_p$, charge $-e$) in front of a mirror consisting of electrons and heavy ions (mass $20m_p$, charge $-e$)} in a simulation box with $2048\times3072$ cells and a size of approximately $10\times20$\,$\mu$m. \Felix{The mirror is placed at $x=0$ and the thin plasma layer at $x=0.4\lambda_0$, to account for imperfect placement of the layer not exactly at $\lambda_0/2$}. We checked \Felix{that with a tolerance of $\pm\lambda_0/4$ changes} in the layer's initial position, possibly due to fabrication inaccuracy, do not significantly affect CSWA, \Felix{since the standing wave's nodes capture all electrons within half a wavelength, establishing an efficient auto-stabilization of CSWA}. We choose the initial conditions such that at the simulation start the center of the chirped laser pulse is placed approximately $5$ times its spatial width in front of the layer.
\begin{figure}[t]\centering
 \includegraphics[width=\linewidth]{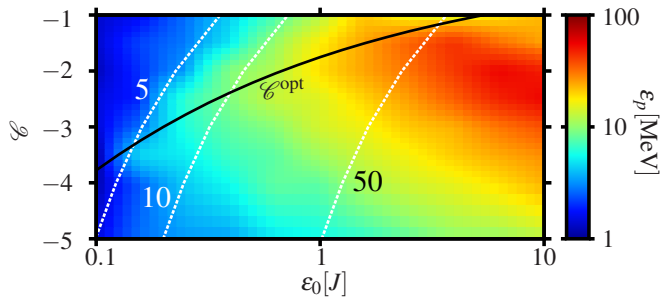}
 \caption{(color online) Maximum proton energy $\varepsilon_{p}$ (energy above which 1$\%$ of all protons lie) as a function of the pulse chirp $\chirp$ and the laser energy $\varepsilon_0$. All parameters as in Fig.~\ref{fig:PhaseSpace} except $\Delta\omega_0=0.3\omega_0$. Black line: Optimum chirp from Eq.~(\ref{Eq:OptimalChirp}). White dashed lines and adjacent numbers: Maximum proton energy from Eq.~(\ref{Eq:OptimalMomentum}) in MeV.}
 \label{fig:ParameterScan}
\end{figure}
First the incident laser radiation is reflected by the layer, causing a minor internal shift of the layer's electrons and a subsequent proton acceleration in positive $x$-direction. Surpassing the threshold intensity ($t \approx 12$~fs), the incident radiation starts to penetrate the layer, forms a standing wave in front of the mirror and captures the electrons within less than $3$~fs. They follow the field node and the generated quasi-static longitudinal field starts to accelerate protons in the negative $x$-direction ($t \approx 15-30$~fs) with the electron and proton layers significantly compressed and stabilized by the laser's ponderomotive force (s.~Fig.~\ref{fig:PhaseSpace}~(b)). The acceleration continues after the laser radiation has decoupled from the electron layer with in this second stage the standing wave being provided by a radiation mode locked between the layer and the mirror ($t \approx 30-80$~fs), accelerating the protons to energies beyond $140$ MeV. \Felix{The protons' distribution is clearly peaked at high energies and in position, proving the strong suppression of plasma instabilities in CSWA even at late times (s.~Fig.~\ref{fig:PhaseSpace}~(c),(f),(j)). All high-energy protons propagate within a narrow cone around the negative $x$-axis, indicating a low beam divergence. Due to the tremendous total accelerated charge, energy-selecting protons, as is customarily done \cite{Busold_etal_2014,Agosteo_etal_2014}, yields $4\times10^9$ ($2\times10^9$) protons in a $\pm2.5$~MeV  window around $85$ MeV ($110$ MeV) without any further optimization (s.~Fig.~\ref{fig:PhaseSpace}~(j)). As indicated above, however, an increased monochromaticity is also intrinsically achievable at the cost of reduced ion energies.}

We confirmed that for laser spot sizes $d_0\gtrsim7\,\mu$m, the protons' spectral properties from a 2D simulation were in good agreement with those from a 1D simulation. This observation further indicates that CSWA is well described by a 1D approximation. For smaller spot sizes in 2D simulations the protons' monochromaticity is reduced while their maximal energies are still in good agreement with those from 1D simulations. Then, to highlight CSWA's wide applicability range and tunability through the chirp, we performed a parameter scan of the maximal proton energies with 1D simulations. We assumed a reduced spot size of $d_0=5\,\mu$m and the plasma layer to be initially placed at $x=\lambda_0/2$, with all other parameters unchanged compared to the previous example. We then varied the total pulse energy $\varepsilon_0$ and the chirp $\chirp$, (s.~Fig.~\ref{fig:ParameterScan}). Comparing the results to $\chirp^\text{opt}$ from Eq.~(\ref{Eq:OptimalChirp}), for small pulse energies we find very good agreement with the theory. For large pulse energies, on the other hand, the chirp of optimal proton acceleration becomes larger than $\chirp^\text{opt}$, probably due to $\chirp^\text{opt}$ reducing to small values and simultaneously the second stage of the acceleration becoming dominant. The displayed proton energy thresholds (dashed lines), obtained from Eq.~(\ref{Eq:OptimalMomentum}), are also well reproduced for small chirp values, while for $\chirp>\chirp^\text{opt}$ the acceleration's efficiency reduces since the protons catch up with the electron layer, preventing them from achieving high energies. Thus, CSWA is demonstrated to yield proton energies of order $\varepsilon_{p} \approx p_{p}^2/2m_p \sim 100$ MeV for pulse energies $\varepsilon_0\sim10$ J \Felix{while simultaneously offering control over their phase space distribution.} Consequently, Eq.~(\ref{Eq:OptimalMomentum}) indicates that at $\varepsilon_0\sim100$ J, CSWA could admit \Felix{controlling} dense and collimated proton beams of up to $\varepsilon_{p} \sim 1$ GeV.

In summary, we presented and analyzed a novel approach to laser ion acceleration, efficiently converting a laser's transverse into an accelerating field and insusceptible to the formation of plasma instabilities. We demonstrated its feasibility over a wide range of realistic parameters, and presented a scaling law indicating the feasibility of GeV-level ion acceleration. We highlighted a unique tunability, enabling a tradeoff between the number and energies of accelerated particles.

AG came up with the original idea for the acceleration mechanism. FM developed the analytical tools to model the chirped pulse and the acceleration mechanism. FM and AG conceptually designed the simulations and FM performed them. All authors analyzed the results and wrote the manuscript.

The authors acknowledge valuable discussions with C.-G.~Wahlstr\"om, O.~Lundh and J.~Magnusson, technical support by the PICADOR development team (especially by S.~Bastrakov) as well as financial support by the Wallenberg Foundation within the grant "Plasma based compact ion sources" (PLIONA). The simulations were performed on resources provided by the Swedish National Infrastructure for Computing (SNIC).


\end{document}